# The Itinerancy and Interactions of the Linear Strings of Holes in Copper-Oxide Superconductors.

Moshe Dayan

Department of Physics*, Ben-Gurion university,
Beer-Sheva 84105, Israel.

* Retired from Ben-Gurion University.

# The Itinerancy and Interactions of the Linear Strings of Holes in Copper-Oxide Superconductors.

Moshe Dayan

Department of Physics*, Ben-Gurion university, Beer-Sheva 84105, Israel.

## Abstract

Here I present a new model for the itinerancy of the strings of holes in the Cuprates HTSC. The model assumes various scenarios with respect to the order of the holes hopping and evaluates the weighting parameters for the different scenarios. The new model still results in the aggregation of holes into strings, but yields a spectral distribution for the itinerancy rates of the strings. From this distribution I infer a spectral distribution for the magnetic interaction between the strings, which suggests also a spectral distribution for the pseudogap parameter, and some relevant experimental functions. Apart from these distributions, the basic assumptions of former relevant theories remain intact. Such assumptions are the existence of the anti-ferromagnetic phases A and B, the basic structure of the pseudogap ground state, the excitation operators, and the field. The ground state and the field are basically divided into two bands, the gapless low energy band, and the high energy band. Due to the wide distributions, the bands may be partially overlapped.

## 1. INTRODUCTION.

More than three decades have passed since the first experimental discovery of high temperature superconductivity (HTSC) in Cuprates. Since then, there has been extensive research on the subject, both experimental and theoretical. The experimental research has accumulated large and important volume of data, which brought important characterizations and insight [1,2,3]. The theoretical research, however, has been less successful, since it has not yet provided a consensual theory of the phenomenon. This is so despite of many various attempts, starting soon after the experimental discovery. Already in 1987 Mott wrote: "There are as many theories as theorists", and he continued in 1991 that: "It is almost still the case now" [4]. The theory that has accepted the support of some researchers, but that has not become consensual, is the RVB theory of P. W. Anderson, that was first proposed in 1987 [5].

The present author started a theoretical study of some aspects of the subject roughly a decade ago, which has yielded some preliminary publications [6-9]. Only eight years ago, these studies have been matured to have yielded a significant paper whose title is: "The Origin of the Pseudogap in Underdoped HTSC"[10,11]. The basic features of the "Origin" are almost self-suggestive. The most basic presumptions of the "Origin" are the following. The t-J Hamiltonian is assumed to be the proper Hamiltonian to deal with the high temperature superconductors (HTSC) cuprates. The t-J Hamiltonian has been used by various works [12-14]. It is given by Eq. (1) in [11], where it was divided into its hopping part - $H_t$, and its magnetic part - $H_J$, namely:

$$H_{tJ} = H_t + H_J.$$

In the un-doped anti-ferromagnetic parent materials, every spin is surrounded by other four anti-parallel spins that results in reducing the average magnetic energy by $J$ per spin. Suppose now that a certain amount of holes, which makes a fraction $\delta$ of the total number of unit cells, is doped into the parent material to produce under-doped HTSC cuprate. Let us also assume that each of these holes is segregated, and it is not the nearest neighbor to any other hole. Then, the magnetic energy per one plane of $N^2$ unit cells, is $-N^2 J(1-2\delta)$. Thus, the magnetic energy is raised by $2J$ per hole. If, on the other hand, all the holes are agglomerated into one squared area, and we neglect the energy of the edges, then the magnetic energy increase per hole is only $J$, one half of the energy of the segregated holes. Obviously, when Coulomb interaction and holes hopping are not considered, it is energetically favorable to agglomerate the holes. So far, we have not considered hopping energy and its consequences, which happen to be very important in HTSC Cuprates. When $H_t$ is turned on, segregated holes become even less acceptable, because their movements would destroy anti-ferromagnetism, not only at their close neighborhood, but also across their former tracks. As about the big single conglomerate, it is also unacceptable because it cannot preserve its form after the holes start hopping. Experiments suggest strongly that the holes agglomeration should be in the form of straight strings, in the vertical and the horizontal directions, as in checkerboard geometry. When a large number of holes are aggregated to form a straight string, their magnetic energy increase is $\frac{3}{2}J$ per hole, an intermediate between the increase of the segregated holes, $2J$ per hole, and of the big single conglomerate, which is $J$ per hole. The big advantage of the said checkerboard linear aggregation of holes is their itinerancy, and their ability to enable the pseudogap state, and superconductivity. The itinerancy of the linear aggregation of holes has been demonstrated in [11]. However, since some steps of that procedure are questionable, the subject of the itinerancy will be re-evaluated, and consequently, will be better established in the present paper.

The agglomeration of the holes into itinerant columns and rows (in a checkerboard geometry), each separated by two phase-inverted anti-ferromagnetic areas of spins, leads immediately to important consequential agreement with experiment. Simple considerations of how to contrast the reciprocal space of the described physical picture are given in the "Origin", considerations which led directly to its Fig. 1a. The agreement with experiment is demonstrated by Fig. 1b [11]. Fig. 1a in the "Origin" depicts the boundaries of areas in the Brillouin zone (BZ) that enclose the ground state of the spins and holes in the two dimensional $CuO_2$ planes. It is often referred to as the "underlying Fermi surface" (UFS). Other theoretical models have obtained such a rough agreement with experiment only after elaborate calculations, using **unjustifiable adjustable parameters**. One such parameter is the second nearest neighbor hopping parameter- $t'$, which in general is adjusted to be as a significant fraction of the first nearest neighbor hopping parameter- $t$ [14, 28, 29]. Recalling that the second nearest neighbor is located in the nodal directions $\pm(1,1)$ and $\pm(1,-1)$, implying that such an adjustment is unacceptable. This is so because the strong directionality of the $d_{x^2-y^2}$ lobes of the Copper, and the $p_x$, $p_y$ lobes of the oxygens suggest the strong exponential diminishing of hopping in the nodal directions. Contrary to this unacceptably adjustment of parameters, the BZ in Fig. 1a is an unavoidable consequence of the itinerant columns and rows of holes in the two dimensional $CuO_2$ planes, as shown in [11]. Besides, without the assumption of holes aggregation, hopping of holes lead to losing local anti-ferromagnetic order, contrary to Neutron scattering measurements, where the elastic incommensurate peaks appear at $\frac{\pi}{a}(1\pm2\delta,1)$ and $\frac{\pi}{a}(1,1\pm2\delta)$. They appear as satellites around the wave-number of the anti-ferromagnetic order- $\frac{\pi}{a}(1,1)$, indicating only wave modulations around that order, without its destructing, which is exactly the result of the "Origin".

    The main features of the UFS of Fig. 1a in the "Origin" are the following: The holes states, which are enclosed by the shaded areas, are located in the anti-nodal directions, around the boundaries of any two adjacent BZ. They are located within rectangles of $(\pi/2a)$ by $(2\pi\delta/a)$. Obviously, the boundaries of these shaded areas are nested, since the whole physics is semi one dimensional. However, there are differences between these two kinds of nesting, as has been explained in the "Origin". The nesting in the string direction produces the 4x4 CDW, whereas the nesting parallel to the string direction produces the incommensurate Neutron Scattering peaks [11]. The width- $(\pi/2a)$, in the strings direction, is a direct consequence of the checkerboard geometry, as explained in the "Origin". If the physics was in accordance with strip models [15-17], then this width should be doubled, which would be against STM and ARPES experiments [11]. The width- $(\pi/2a)$, in the strings direction, is bordered by two nested straight lines at $\pm(\pi/4a)$. Usually, such situations create some irregularities in the spectrum of the states at these wave-numbers, which could be reflected in electronic and ionic polarizations. The states with the wave-numbers- $k=\pm\frac{\pi}{4a}$, in the string directions, are scattered and interfered by each other to create a self-sustained field of $k=\frac{\pi}{2a}$. This is a charge density wave electronic field, which

may produce an ionic field, by means of the electron-ion interaction. The wavenumber width of $(\pi/2a)$ fits a modulation by a period of four lattice units in the real space, which makes the field self sustained. Such modulations have been measured by STM and reported in several papers [18-20]. Qualitatively speaking, such perception has been described in the "Origin", although no quantitative analysis has been given so far to the energy spectrum of the states in the string directions.

Nesting in directions that are transverse to the strings creates interferences between string states of wave-vectors $k$ and $\bar{k}$, where $k - \bar{k} = sign(k) 2k_F = sign(k) \frac{2\delta\pi}{a}$. This, as is usual in such mixings, creates an energy gap that separates the particle-like spectrum from the anti-particle-like spectrum. The parameters of the above mixing may be of the same sign or reversed in sign. It turns out that, besides the energy-gap, there are energy shifts of the above mentioned spectra. For reversed sign parameters, both the particle-like and the anti-particle-like spectra, shift towards the Fermi level, and bridge the gap. Thus, converting the gap into a pseudogap. The experimental fact that the HTSC Cuprates are conducting, and even superconducting, stems from these shifts of the energy scales. These conclusions have been reached already in the "Origin". The present analysis obtains large imaginary parts for both the itinerant states, and the pseudo gaps, which probably has important implications on the density of states and the transport properties of the pseudogap state.

The above mentioned mixing of every two states whose wave-number difference is $\pm 2k_F$, produces SDW of the same wave-vectors normal to the string directions. These results are in a good agreement with elastic Neutron scattering experiments, where incommensurate peaks around the anti-ferromagnetic wave-number, at $\frac{\pi}{a}(1 \pm 2\delta, 1)$ and $\frac{\pi}{a}(1, 1 \pm 2\delta)$, have been reported by many authors [21-25]. The fact that these peaks have been observed in elastic Neutron scattering experiments is additional evidence that we have made the proper choice of the sign of mixing of the $k$ and $\bar{k}$ states.

In this introduction, I have used concepts and perceptions that are typical to systems of Fermions. The reader could rightly question this, because usually excitations of collections of Fermions, like Plasmons and Magnons, are Boson-like.

Here the strings of holes are treated like Fermions because a crystal cell of the $CuO_2$ plane can have either one hole or one spin. This suggests the relations: $C_j |0> = 0$, $C_j C_j^\dagger |0> = 1$, and $C_j^\dagger C_j^\dagger |0> = 0$, where $|0>$ is the vacuum anti-ferromagnetic state, and $C_j^\dagger$ creates a string of holes in the j column (row) as defined in the "Origin". In addition one has the freedom of choosing phase so that $\{C_j^\dagger C_i^\dagger + C_i^\dagger C_j^\dagger\}|0> = 0$. These relations are the Fermi anti-commutation relations. Despite the above relations, we shall see during the analysis of the present paper that the strings of holes are far from satisfying the basic requirements of Landau's Fermi systems. This is so because the dispersion of the self-energies of the "quasi-string-particles" is much smaller than the magnetic interactions between them. This feature, together with the broad spectrum of their itinerancy, result in inverse life-times that are of the same order as the excitation energies. However, the strict Landau criteria for Fermi systems are violated also in other systems, such as for example highly disordered metallic system where the electron conductivity is affected by Coulomb blockade or Coulomb gap. This suggests that one should be less restrictive in using Fermi system concepts in dealing with systems that are not strictly Landau's Fermi systems.

The main purpose of the present paper is to re-establish the theory of the linear strings of holes in Cuprate HTSC. This will be done by reassuring the basic model, together with the correction of some faults. The two main parts of this re-establishment are: 1) The re-evaluation of the itinerancy of the string states that are arranged in a checkerboard geometry, its deduced dispersion and life times. 2) The re-evaluation of the magnetic interactions and their consequential pseudogap order parameter. 3) The implications to some properties of the pseudogap state.

## 2. MODELING THE ITENERANCY OF THE LINEAR STRINGS OF HOLES.

In the "Origin" the holes were assumed to conglomerate into linear strings which make rows and columns, and which move by means of the application of $H_t$ on the neighboring spins repeatedly and continuously, one by one. This is done formally by applying the time development operator on the string. During the process the magnetic energy is increased by $J$, but comes back to its original value after the whole string is moved by one crystal unit. This energy restoration results from the magnetic energy restoration of the string, and from the fact that on both sides of the string there are two perfect anti-ferromagnetic regions, as before the string movement. Here I shall re-evaluate this process, discuss its faults, and suggest an alternative.

The t-J Hamiltonian is given by

$$H_{t-J} = -\tau \sum_{<ij>} (a_{is}^\dagger a_{js} + hc) + J \sum_{<ij>} (S_i \cdot S_j - \frac{1}{4} n_i n_j)$$
$$= H_t + H_J. \tag{1}$$

The various parameters in the equation were defined in the "Origin". Notice that the t-J Hamiltonian does not include Coulomb interaction. The time development operator is given by

$$U(t,-\infty) = \sum_{n=0}^{\infty} U_n = \sum_{n=0}^{\infty} (-i)^n \int_{-\infty}^{t} dt_n \int_{-\infty}^{t_n} dt_{n-1} \cdots$$
$$\times \int_{-\infty}^{t_2} dt_1 H_t(t_n)...H_t(t_1) e^{-\varepsilon(|t_1|+...|t_n|)} \tag{2}$$

The time development operator operates adiabatically, by means of the perturbation Hamiltonian, on states of the unperturbed Hamiltonian, to result in the approximate solutions of the total Hamiltonian. In the usual perturbation problem the unperturbed Hamiltonian is soluble, and the perturbation Hamiltonian presents only a small perturbation, so that the application of a few lower orders of $U_n$ is sufficient. In the present problem, the time development operator was used in an unusual manner. The magnetic Hamiltonian $H_J$ was taken to be the unperturbed Hamiltonian, because its eigenvalue is known for any configuration of spins. Given the small amounts of doping, in the under-doping regime, the absolute value of the magnetic energies is still much larger than the kinetic energies due to the itinerancy of the strings. Nevertheless, the kind of application of the perturbation theory in the "Origin" was unusual, because the kinetic parameter $\tau$ is larger than the magnetic parameter $J$.

The magnetic coupling is given by $J = \frac{4\tau^2}{U}$, where it is usually assumed that $U > 4\tau$, so that $J < \tau$. In the literature the ratio $(\tau/j)$ was reported to be between 2.5 and 4.0 [12,14,16], and it is reasonable to assume that it is roughly equal to 3.0. The application of $U_n$ on $C_j^\dagger$ produces the pre-factor: $(\tau/j)^n >> 1$, which suggests that $U_n$ of the largest order dominate, which means that hopping of only part of the spins to the next column is insignificant.

There is a major problem with the above described application of the time development operator (beside additional problems). When $U_N$ is applied, where $N$ is the number of cells along the string directions, the total time to move the string one step becomes semi-infinite. This is so because the one step hopping time of each spin is at least of order $\tau^{-1}$, so that the total time necessary to move the whole string is at least of order $N\tau^{-1}$. In the following I present a new model for the itinerancy of $C_j^\dagger$, which suggests that the hopping time for each spin is even larger than $\tau^{-1}$. Anyway, this makes the states $C_q^\dagger$ stationary and dispersion-less. The dispersion given by Eq. (18) in the "Origin" is erroneous because it does not take into account the above time consideration. ARPES data suggest that the dispersion of $C_q^\dagger$ is quite small [26-30], but it is not as small as $\tau/N$. In the following analysis I present a new model for the itinerancy of $C_j^\dagger$. The main innovation of the new model is that it takes into account the rows of holes while applying $U_n$ on the columns of holes, and wise versa. Fig.1 depicts an area of spins and holes that includes parts of two columns and three rows. Each part of a column that is enclosed between two rows will be referred to as a segment.

Fig. 1. Two column segments (and their surroundings) after the first mutual hops that start the movement of the whole columns. The right column and the rows are in different hopping sequences than the left column. The signs indicate spin projections on the z-axis. Cells with holes are colored blue.

Since there are $\frac{1}{2}\delta N$ rows along each column, the **average** number of holes in each segment is $2/\delta$, which equals 16 for a typical under doping parameter of $\delta = 0.125$. If one excludes the holes that are shared by the rows of holes, then this number is: $l = 15$. The numbers of holes in a segment in the figure is arbitrarily chosen to be ten (not including the hole in the row). The new model for the strings itinerancy applies $H_t(t)$ as a sum, where the different terms of the sum operate simultaneously on different segments. This is depicted in Fig. (1), where one sees three spins, in three different segments of the same (left) column of holes that have just hopped to the lowest position of their segment.

The other innovation of the new model is the starting of the adiabatic application of the time development operator at a finite time, $t = 0$, which is necessary in order to avoid the infinite times in Eq. (2). Suppose that one tries to calculate the time development operator by calculating a typical integral that hops one spin from the edge of a segment in the column $j+1$, into the edge of the same segment in the column $j$,

$$I(t_2) = -i\int_0^{t_2} dt_1 H_t(t_1) C_j^\dagger = i\tau \int_0^{t_2} dt_1 a_{1,j,s_1}^\dagger(t_1) a_{1,j+1,s_1}(t_1) C_j^\dagger. \tag{3}$$

In Eq. (3) the time dependence of the creation and annihilation operators is

$$a_{1,j,s_1}^\dagger(t_i) = a_{1,j,s_1}^\dagger \exp(-iJ\frac{n_{1,j}}{2}t_1). \tag{4a}$$

$$a_{1,j+1,s_1}(t_i) = a_{1,j+1,s_1} \exp(iJ\frac{n_{1,j+1}}{2}t_1). \tag{4b}$$

In equations (4) $n_{1,j+1}$, and $n_{1,j}$ are the nearest neighbors of opposite spins before, and after the hopping, respectively. Carrying on the integration in (3) yields

$$I(t_2) = \frac{2\tau}{J\hat{n}}[\exp(iJ\frac{\hat{n}}{2}t_2) - 1]a_{1,j,s_1}^\dagger a_{1,j+1,s_1} C_j^\dagger, \tag{5}$$

with $\hat{n} = n_{1,j+1} - n_{1,j}$. The unity that results from the lower time limit of the integral indicates that the application of $H_t(t_1)$ has not been adiabatic, since it yielded a finite result even at $t = 0$. To eliminate this difficulty, I multiply each $H_t(t_i)$ by $(1-e^{-2\tau t_i})$. Now, $H_t(0) = 0$, and $H_t(t_i)$ assumes its regular value as in Eq. (1), at $t_i \gg \tau^{-1}$. The adiabatically applied $H_t(t)$ becomes

$$H_t(t) = \tau(e^{-2\tau t} - 1) \sum_{<ij>} a_{is}^\dagger(t) a_{js}(t). \tag{6}$$

The adiabatic application of Eq. (6) is convenient, because when we perform the integrations of the time development operator, we obtain zero for the lower limit. For the upper limit we may neglect the first term with the factor $e^{-2\tau t}$, and have only the second (ordinary) term. Thus, with this kind of adiabatic application, the result of Eq.(5) is replaced by

$$I(t_2) = \frac{2\tau}{J\hat{n}} \exp(iJ\frac{\hat{n}}{2}t_2) a_{1,j,s_1}^\dagger a_{1,j+1,s_1} C_j^\dagger \tag{7}$$

In the new model for the strings itinerancy, $H_t(t)$ is applied as a sum, where each term of the sum operates on a different segment. Each such term then continuous the time development by the hopping of other holes of the segment. At the end, the results of all segments are multiplied to yield the whole string. Here, however, I write the results of only one representative segment. This mode of simultaneous operation applies to rows as well as to columns. Here we examine a hopping sequence in which the first spin to hop in a segment is located at the edge of the segment, a choice that makes the exponential in Eq. (7) equals- $\exp(i\frac{J}{2}t_2)$. This exponential in (7) expresses an intermediate energy excitation. To incorporate a built-in mechanism that takes into account the finite life-time of an intermediate energy excitation, let us add to the energy parameter $J$, in equations (4), the inverse of its life-time parameter- $\varepsilon^{-1}$. Consequently, $J$ in Eqs. (4a) and (4b) should be replaced by: $\tilde{J} = J + i\varepsilon$. Now everything is set for the application of the time evolution operator according to the scenario whose beginning is depicted in Fig. 1. With the replacement of $J$ by $\tilde{J}$, setting $\hat{n} = 1$, Eq. (7) is replaced by

$$I(t_2)C_j^\dagger = \frac{2\tau}{\tilde{J}}\exp(i\frac{\tilde{J}}{2}t_2)a_{1,j,s_1}^\dagger a_{1,j+1,s_1}C_j^\dagger. \tag{8}$$

The next step is to apply $H_t(t_2)$ on $I(t_2)C_j^\dagger$, and perform the $dt_2$ integration. The only time dependence of the integrand is the one seen in Eq. (8), because $\hat{n}=0$ for $a_{i,j,s_i}^\dagger a_{i,j+1,s_i}$ with any row index $i$ that is different from the edges, namely for $1<i<l$. Consequently, after setting $n=l$, the result after the $(l-1)t\mathrm{h}$ integration is

$$I_l(t_l) = (\frac{2\tau}{\tilde{J}})^{l-1}\exp(i\frac{\tilde{J}}{2}t_l)a_{l,j+1,s_l}^\dagger a_{l,j,s_l}C_{j+1}^\dagger. \tag{9}$$

On performing the last integration we note that the time dependence of $H_t(t_l)$ cancels the one in Eq. (9), so that the integrand of the $dt_l$ integration is time independent. This would yield a result that is linear in time. This is disturbing since it suggests that the final time parameter in $U_l(t,0)C_j^\dagger$ is limitless- the more you wait- the larger is the final result.

$$U_l(t,0)C_j^\dagger = i\tau t(\frac{2\tau}{\tilde{J}})^{l-1}C_{j+1}^\dagger = \frac{i\tilde{J}t}{2}(\frac{2\tau}{\tilde{J}})^l C_{j+1}^\dagger \tag{10}$$

In the present problem, the time cannot be limitless, because $H_t(t)$ continuous to act as a perturbation even after the hopping of the last spin to the $j-th$ column. The index of $U_n(t,0)$ is not limited to $n=l$, as seen from Eq.(2). This suggests that the time $t$ in (10) should be replaced by some fixed time average- $\bar{t}$. Moreover, the exponential factor in (9) is $\exp(i\frac{\tilde{J}}{2}t_l)=\exp(i\frac{J}{2}t_l-\frac{\varepsilon}{2}t_l)$. The decay factor $\exp(-\frac{\varepsilon t_l}{2})$ reduces the large factor $(\frac{2\tau}{\tilde{J}})^{l-1}$, and therefore renormalizes the time evolution of $C_j^\dagger$. The removal of this renormalizing factor from (10) is unacceptable, and suggests the reevaluation of our calculation procedure, which is done in the following.

We expect a time decay factor because we have implemented a time decay mechanism to limit the time duration of the extra energy excitation that is caused by "breaking" a column. The exponential decay factor is completely missing from (10), because (10) represents a final result of an unbroken holes column. This is unacceptable because, although the final result of (10) is unbroken column, it has been obtained during intermediate time with the energy excitation of a broken column. The time parameter of the implicit decay factor- $\exp(-\frac{\varepsilon t_l}{2})$ in (9) includes the time that has passed to reach (10), which implies the need for renormalizing (10) by the same decay factor. I have considered two ways to correct this problem. The first is to apply the time dependence of the last $H_t(t_l)$ by means of equations (4a,b) with $\exp(-\frac{iJt_l}{2})$, instead of $\exp(-\frac{i\tilde{J}t_l}{2})$. This would yield a result that includes the exponential time decay factor- $\exp(-\frac{\varepsilon t}{2})$, but the choice of this time dependence of $H_t(t_l)$ is doubtful. The second way to obtain the time decay factor in (10), is based on evaluating the times- $t_m$, which appear in the integrals of the time development function. This is discussed below.

Generally speaking, the perturbation by $H_t$ causes a mixing between $C_j^\dagger$, and $C_{j\pm1}^\dagger$, for most of the time. The movement of strings of holes from $C_j^\dagger$ to $C_{j\pm1}^\dagger$, is not sudden as for single spins. It is gradual and takes long intermediate times. To see this let us write $I_l(t_m,0)C_j^\dagger$, as an intermediate time development of $U_l(t,0)C_j^\dagger$, for $1 < m \leq l$. It is equivalent to (9), but with (m-1) integrations. After substituting $\tilde{J}^{-1} = |\tilde{J}|^{-1} e^{-iarctg(\varepsilon/J)}$ we get,

$$I_l(t_m,0)C_j^\dagger = (\frac{2\tau}{|\tilde{J}|})^{m-1} \exp\{i[\frac{J}{2}t_m - arctg(\frac{\varepsilon}{J})(m-1)] - \frac{\varepsilon}{2}t_m\}C_{m-1,j+1}^\dagger C_{l-m+1,j}^\dagger. \quad (11a)$$

In (11a) $C^\dagger_{m-1,j+1} C^\dagger_{l-m+1,j}$ denotes a "broken column", where (*m-1*) holes have been transferred to the column (j+1), and (l-m+1) still remain in the column *j*. For *m=l* we get

$$I_l(t_l,0)C^\dagger_j = (\frac{2\tau}{|\tilde{J}|})^{l-1} \exp\{i[\frac{J}{2}t_l - arctg(\frac{\varepsilon}{J})(l-1)] - \frac{\varepsilon}{2}t_l\}C^\dagger_{l-1,j+1}C^\dagger_{1,j}. \qquad (11b)$$

Note that the phase of the exponent includes a shift of $[-arctg(\frac{\varepsilon}{J})(l-1)]$. Since *l* is an index of the holes along the string, the imaginary part of the exponent describes a wave along the string direction, a wave of the spins that have hopped from the column j+1 to the column j. The absolute value of its wavenumber is $arctg(\frac{\varepsilon}{J})/a$, and its phase velocity is $v = aJ/2arctg(\varepsilon/J)$. Thus, we get

$$t_l^0 = \frac{(l-1)a}{v} = \frac{2(l-1)arctg(\varepsilon/J)}{J}. \qquad (12)$$

In (12) $t_l^0$ has the superscript zero to distinguish it from the last integration time variable $t_l$. Now we set the time dependence of $H_t(t_l)$ to be- $\exp[-i\frac{\tilde{J}(t_l - t_l^0)}{2}]$, which results in

$$U_l(t,0)C^\dagger_j = i\tau t(\frac{2\tau}{|\tilde{J}|})^{l-1}\exp[-\frac{\varepsilon}{J}(l-1)arctg(\varepsilon/J)]C^\dagger_{j+1}. \qquad (13)$$

Equation (13) is similar to (10) except for time decay factor- $\exp(-\frac{\varepsilon t_l^0}{2})$. This decay factor makes an important correction to the weighting factor of the described process. The calculation of (13) implies a rule which one should keep in general: With the application of any new $H_t(t_m)$ which implies a change of the intermediate energy excitation, the zero of its time dependence should coincides with its application time, as demonstrated in the calculation of (13). The rate for transporting all the holes from a column (or a row) to its adjacent column (or row) by the described process is approximately

$$r_1 \cong \frac{J}{2(l-1)arctg(\varepsilon/J)}. \tag{14}$$

Now, we begin to get better insight into this different kind of perturbation, where $H_t$ perturbs $H_J$, and where $3 \cong \tau/J$. The application of $U_l(t_m,0)C_j^\dagger$, in a sequential manner starting from an edge of a segment, produces the pre-factor $(2\tau/|J|)^{m-1}$ in (11). This pre-factor is much larger than unity, and becomes exponentially larger with the number of spins hops. However, this large pre-factor, is only one of the factors that determine the relative weighting of the various possibilities of calculating $U_l(t_m,0)C_j^\dagger$. Another one is the time decay factor $\exp(-\varepsilon t_m/2)$ in (11). Still another one is the probability of the occurrence of the sequence of the spin hopping process.

As stated before, we have chosen a particular model in which the first hole to hop from a column is the nearest neighbor to a row of holes, and the next holes hop in a sequence. Some readers may think that such a particular choice is artificial, and wonder how the system makes its choices anyway? Why, for example, different terms of the sum of $H_t(t)$ do not operate on different sites of the same segment? The answer is that such a mode of operation would create a sum of terms for one segment. When the results for all the segments are finally multiplied, the product cannot represent one string, even not a "broken" one. Instead, it is a sum of terms, where each is a product of holes from different segments, and different terms represent different holes. Results of this type are contradictory to experiment. Thus, if one wants to calculate $U_n(t,0)C_j^\dagger = (-i)^n \int_0^t dt_n H_t(t_n) .... \int_0^{t_2} dt_1 H_t(t_1) C_j^\dagger$, where $H_t(t_m)$ is given as a sum of hops to neighboring positions, then every term of the sum should be applied on a different segment. The following sequential applications of the $H_t(t_m)$'s should keep each time sequence within each segment. The separation by rows (of holes) between segments of column (of holes) is a natural separation to break the sequential application of $H_t(t_m)$. It does not mean that the scenario that has been described in the former sections (which hereafter is referred to as scenario 1) is the only possible one. In the following we examine other scenarios.

Another sort of perturbation which is examined below assumes that the first spin to hop to the string of holes is not located at an edge of the segment, but somewhere along the segment (hereafter this scenario is called scenario 2). Suppose that the first spin to hop to $C_j^\dagger$ is located m rows away from the nearest row of holes that borders the segment, namely $1 < m \leq l/2$. We assume a sequential hopping of the spins from the column (j+1) into the column j, starting from the spin at m and covering all the spins except for the two at the edges. Here "sequential hopping" means that we keep the order of spin hopping so that there are exactly two break points before the hopping of the last edge spins. This means that $(l-2)$ spins should hop, and $(l-2)$ time integrations should be performed on integrands with the time dependence of $\exp(i\tilde{J}t_i)$, which should yield the pre-factor

$$I_l(t_{l-1},0)C_j^\dagger = (\frac{\tau}{|\tilde{J}|})^{l-2} \exp\{i[Jt_{l-1} - (l-2)\cdot \text{arctg}(\varepsilon/J)] - \varepsilon t_{l-1}\}C_{2,j}^\dagger C_{l-2,j+1}^\dagger \quad (15a)$$

$$t_{l-1}^0 = \frac{(l-2)}{J}arctg(\frac{\varepsilon}{J}) \quad (15b)$$

The next integration is $\int_0^{t_l} dt_{l-1} \exp(i\tilde{J}t_{l-1})H_t(t_{l-1})$, where the time dependence of $H_t(t_{l-1})$ is $\exp[-i\frac{\tilde{J}}{2}(t_{l-1} - t_{l-1}^0)]$. This yields

$$I_l(t_l,0)C_j^\dagger = 2(\frac{\tau}{|\tilde{J}|})^{l-1} \exp\{\frac{i}{2}[\tilde{J}t_l - l\cdot \text{arctg}(\varepsilon/J)] - \frac{\varepsilon(l-2)}{2J}arctg(\varepsilon/J)\}C_{1,j}^\dagger C_{l-1,j+1}^\dagger$$
(16a)

$$t_l^0 = \frac{l}{J}arctg(\frac{\varepsilon}{J}) \quad (16b)$$

The last integration yields

$$U_l(0,t)C_j^\dagger = i\tau t 2(\frac{\tau}{|\tilde{J}|})^{l-1} \exp[-\frac{\varepsilon}{J}(l-1)\cdot \text{arctg}(\varepsilon/J)]C_{j+1}^\dagger \tag{17a}$$

The rate is the inverse of $t_l^0$, which is

$$r_2 = \frac{J}{l\cdot arctg(\varepsilon/J)} \tag{17b}$$

It is about twice the rate of scenario 1 because of shorter times.

The two mentioned scenarios for spin hopping, scenarios 1 and 2, are only prototypes. There are many other possible scenarios- $sc_i$; each comes with its probability of occurrence- $p_i$, its weighting factor- $f_i$ , and its transferring rate- $r_i$. Let us now examine scenario 3, which is different from the former two mainly because it produces more than two breaking points along the string in the segment. It is a combination of the former two scenarios, and therefore it produces three breaking points. Scenario 3 starts as scenario 1 by hopping an edge hole, which produces the first break point. The second step hops a hole in the middle of the segment, just as in scenario 2, and produces additional 2 break points. After these two steps we get

$$I(t_3) = 3(\frac{2\tau}{3|\tilde{J}|})^2 \exp\{i[\frac{3}{2}\tilde{J}t_3 - 4arctg(\varepsilon/J)] + \frac{2\varepsilon}{J}arctg(\varepsilon/J)\}. \tag{18}$$

Now, as in scenario 2, we perform $(l-4)$ integrations without changing the number of break points, and get,

$$I(t_{l-1}) = 3(\frac{2\tau}{3|\tilde{J}|})^{l-2} \exp\{i[\frac{3}{2}\tilde{J}t_{l-1} - l\cdot arctg(\varepsilon/J)] + \frac{2\varepsilon}{J}arctg(\varepsilon/J)\}. \tag{19}$$

Now we apply the hopping Hamiltonian twice, in the first step a middle hole hops to eliminate two break points, and in the second step an edge hole hops and eliminates the last break point. As about the time dependence of the relevant $H_t(t_m)$, we apply the rule that has been applied for the former scenarios. Finally we get,

$$U_l(t,0)C_j^\dagger = i\tau t 9 (\frac{2\tau}{3|\tilde{J}|})^{l-1} \exp[-\frac{\varepsilon}{J}(l-2)\cdot arctg(\varepsilon/J)]C_{j+1}^\dagger. \tag{20a}$$

For the rate we get

$$r_3 = (t_l^0)^{-1} = \frac{3J}{2(l+3)arctg(\varepsilon/J)} \tag{20b}$$

A comparison between the three scenarios shows an obvious trend: $r_2 \cong 2r_1$, $r_3 \cong 3r_1$. The rate is proportional to the number of break points, the more break points, the higher is the intermediate energy excitation, and the faster are the hops. Surprisingly, the time decay factor is not strongly dependent on the scenario. It depends only through the ratio $\varepsilon/J$, which according to the uncertainty principle it is assumed to be $\varepsilon/J \geq 1$. On the other hand the pre-factor is largely reduced with increasing the number of the break points. The probabilities- $p_i$, for the three scenarios, have not been evaluated so far.

I have also tried to evaluate scenarios with higher numbers of break points. The scenario with the highest number of break points hops $\mu = Int[(l-1)/2]$ set of holes that none of its members is either an edge hole, or the nearest neighbor to another. We assume that these holes hop, one after the other, to the next column. After this set of hops, the holes between the formerly hopped ones start to hop one after the other. Now every hop decreases the number of break points by 2. We find that the pre-factor of this scenario is $i2t\tau \frac{\mu}{\mu!^2}(\frac{\tau}{\tilde{J}})^{2\mu}$, while the time decay factor is similar to the decay factor of the former scenarios. Thus, the weighting factor is negligible. This last scenario demonstrates the reducing weighting factors of scenarios with large number of break points. We estimate that the limit number of break points for a scenario should be three to four.

Consequently, scenarios 1,2, and 3 are good prototype scenarios, from which other scenarios may be derived. An example of a scenario that might be derived from scenarios 1 and 2 is a scenario that starts as scenario 2 but does not utilize all the possible hops with two break points. Instead, at some early stage it completes the sequence of hops to the close row of holes, and continues from this stage as scenario 1. This is only one example and many other examples may be thought of as combinations between scenario 1 and 2, or 2 and 3, etc. This way, one can think of a continuous spectrum of scenarios, each with substantial probability and weighting factor, and with a rate somewhere between $r_1$ to $r_3$. The time development operator $U_l(t,0)$, is therefore expressed as a sum of all its scenarios' time tracks $U_l^i(t,0)$, so that $U_l(t,0)C_j^\dagger = \sum_i U_l^i(t,0)C_j^\dagger$.

During our discussion of the various scenarios, we have ignored the possibilities of the holes hopping to their initial string positions $C_j^\dagger$. These scenarios may exist, but they only renormalized the string $C_j^\dagger$, and make it dressed by the hopping Hamiltonian $H_t$. They do not change the former analysis in principle, and we have by passed them for the sake of simplicity.

The string operators $C_j^\dagger$, define the string states $C_j^\dagger |0>$, that are not eigen-states of the Hamiltonian, even after transforming them back to the Schrödinger picture. The eigen-states of the Hamiltonian are linear combinations of the string states. Let us denote them by

$$|\phi_k> = \frac{1}{\sqrt{N}}\sum_j B_j(t) e^{-iE_0 t} C_j^\dagger |0> = \frac{1}{\sqrt{N}}\sum_j B(t) e^{-iE_0 t} e^{-ikaj} C_j^\dagger |0> \qquad (21)$$

In (21), the $B(t)$ dependence on time results from transferring holes to neighboring columns (rows) with the rate $r$, whereas the energy $E_0$ represents some zero energy without this transferring. Therefore, the Schrodinger equation for $B_j(t)$ should be

$$i\frac{dB_j(t)}{dt} = E_r(k)B_j(t) = E_0 B_j(t) - rB_{j+1}(t) - r B_{j-1}(t) \qquad (22)$$

Inserting (21) into (22) leads to

$$E_r(k) = E_0 - 2r\cos(ka). \tag{23}$$

Setting arbitrarily our zero energy scale at $E_0$, makes the eigen-state (21) become

$$|\phi_k(t)> = \frac{1}{\sqrt{N}}\sum_j B_j(t)C_j^\dagger |0> = \exp[i2r\cos(ka)t]C_k^\dagger |0> \tag{24}$$

In (24) $C_k^\dagger$ is defined in the forth-coming equation (27). The rate r depends on the order of the spin hopping for a single scenario, while we have a distribution of scenarios $sc_i$ with various weighting factors- $f_i$ and different probabilities of occurrence- $p_i$. Thus, (24) is generalized for multiple scenarios to become

$$|\phi_k(t)> = \frac{1}{\sqrt{N}}\sum_{j,i} B_{j,i}(t)C_j^\dagger |0> = \sum_i B_i \exp[i2r_i\cos(ka)t]C_k^\dagger |0>$$

$$= \sum_i A_i(k,t)C_k^\dagger |0>. \tag{25a}$$

The normalization of $|\phi_k(t)>$ suggests that

$$\sum_i |A_i(k)|^2 = 1. \tag{25b}$$

In (25a,b), $B_i = cp_i f_i = A_i(k,t=0)$, where $c$ is a general normalization constant. The function $A_i$ may be considered the spectral distribution of the states $|\phi_k>$, because of its dependence on the rates of the different scenarios, which in turn determine the energy spectrum. Each $i$- component of $|\phi_k(t)>$ is an eigen-state of the Hamiltonian with the eigen-value $e_i(k) = -2r_i\cos(ka)$. The state $|\phi_k(t)>$ is an eigen-state of the Hamiltonian with the weighted averaged eigen-value $\bar{e}(k) = -\sum_i B_i 2r_i\cos(ka) = -2\bar{r}\cos(ka)$.

We remark again that the string state is a product of all the segmental states. Naturally this product is a weighted sum of terms that are products of many scenarios. We speculate that mixed scenarios terms should be weighted weaker than single scenario terms. This is so because the time development of the rows and the columns are coordinated. Two neighboring column segments that propagate in different rates produce mismatch with the propagation of the row between them. This translates to a reduction in the weighting factor. Consequently, there is a kind of coherence in the sense that terms with same rate scenarios are weighted stronger than terms with mixed scenarios. Thus, (25) that have been derived for one segment may be generalized for the whole string, provided that $A_i(k,t)C_k^\dagger|0>$ is defined as a string distribution function of the string scenario- $sc_i$.

Now let us evaluate the dispersion of our strings. We have already estimated that the three scenarios 1, 2, and 3, define roughly our spectrum range, with scenario 2 in the middle rate. The rate of scenario 2 by (17b) is about 10meV. This makes the Fermi energy only a couple of meV above the bottom of the band. Thus, there is a dispersion of only a couple of mev for a momentum range of $\pi/8a$. This very small dispersion has been measured by ARPES in many investigations [26-30].

We sum up this section by asserting that the strings of holes that are arranged by checkerboard geometry are established both experimentally and theoretically. These strings move gradually by means of combinations of holes hops, that we named- scenarios. They are itinerant, but with kinetic energies of very small dispersion. Different scenarios cause different rate of propagation, causing a spectrum distribution of the kinetic energy for any single state. There is also broadening of the wave-number of the states, since the strings move gradually as broken strings, but this broadening is small in comparison to $k_F$. Although the strings' length is ideally equal to $Na$, there are other lengths that are characteristic of the system. One such length is the length of a segment, namely- $2a\delta^{-1}$. Another one is $4a$, which is the reciprocal of the width of the arms of the Brillouin zone in the anti-nodal direction, as in Fig. 1 of [11].

3. MAGNETIC INTERACTIONS BETWEEN THE STRINGS.

The linear strings that have been analyzed in the last section make a narrow band of itinerant strings. These states make up only the basic ground state from which an ordered state is constructed- the pseudogap ground state [11]. Two features are essential for producing this phase transformation and for giving it its characteristics. One is the semi one-dimensional character of the string states, which produces nesting in the Brillouin zone. The other is the strength of the interactions between strings in comparison with their small itinerant energy. Generally speaking one should consider three sorts of interactions: Coulomb, phonon mediated interactions, and magnetic interactions. Keeping with the line of treatment of the present paper, here we treat only the magnetic interactions.

The analysis of the magnetic interactions between two strings of holes is much more complex than the analysis of their itinerancy that has been made in the former section. This is so because it involves hopping of holes of two strings up to contacts between them and then a continuation of hopping that leads to separation after the contact has been made. Suppose that we examine two segments of columns one at $j$ and the other at $j' = j+3$, and that they move towards each other. Magnetic energy is reduced when holes from $j$ hop to $j+1$ and make contact with holes from $j+3$, which have hoped to $j+2$. This evaluation has to be done at the same time, which means that the two segments have to be examined by the same time evolution operator $U_m(t,0) C_j^\dagger C_{j+3}^\dagger$. Preliminary analysis of this sort posed some difficulties. In the present paper I would rather use the simpler approach that is presented below.

Let us analyze the magnetic interaction between two perfect full linear strings (of $N$ holes each). The interaction in the real space of these two strings, $C_j^\dagger$ and $C_{j'}^\dagger$, is given by []

$$V_{j,j'} = -0.5 J N \delta_{j,j'\pm 1} \tag{26}$$

The interaction energy is negative and none zero only for two closely neighboring columns (rows). The Fourier transform of (26) is defined by means of the strings in the momentum space

$$C_k^\dagger |0> = \frac{1}{\sqrt{N}} \sum_j C_j^\dagger \exp(-ikja) |0> \qquad (27)$$

The interaction in the momentum space is calculated when (26) is sandwiched between two initial states and two final states

$$V(k-k'=q) = <0| \frac{-1}{2N^2} \sum_{j,j'} C_j C_j^\dagger e^{iaj(k-k')} C_{j'} C_{j'}^\dagger e^{-iaj'(k-k')} [NJ(\delta_{j,j'+1} + \delta_{j,j'-1})] |0>$$

$$= -J\cos(qa). \qquad (28)$$

Equations (26) and (28) give the magnetic interactions between two complete perfect strings, in the real and the reciprocal spaces, respectively. Equation (28) has been obtained by considering $C_k^\dagger |0>$ and $C_{k'}^\dagger |0>$ as complete species, without working out the details of their interactions. Thus, its use for interactions between real stings of holes is doubtful. We have seen that real strings are divided into segments, and are broken in a couple of break points (in each segment) during their propagation. If we ignored these fractures, and only normalize the interaction for one segment, then (26) becomes $V_{j,j'} = -0.5Jl\delta_{j,j'\pm 1}$ and (28) becomes $V(q) = -\frac{Jl}{N}\cos(qa)$. However, this normalization does not take into account the complexity of two fractured strings that are first propagating towards each other, toughing in parts, and then propagating away from each other. Suppose that we try to evaluate the interaction in the real space between two segments of the strings $C_j^\dagger$ and $C_{j'}^\dagger$, that are propagating towards each other, and let us assume that at $t=0$, $j'=j+3$. Now, we must consider scenarios that incorporate hopping of holes in both strings, a procedure which complicates the calculations significantly. Such calculations have not been done so far.

Despite the above mentioned difficulties, we assume that the segmental strings approach each other, touch in part of their holes, and then depart away from each other. Thus the magnetic interaction in a segment is

$$V(q) = -\frac{J}{N}\cos(qa)\sum_i \alpha_i(r_i)n_i = -\frac{J}{N}\cos(qa)\bar{n}. \tag{29}$$

For the whole string the interaction is

$$V(q) = -\frac{J}{l}\cos(qa)\sum_i \alpha_i(r_i)n_i = -J\frac{\bar{n}}{l}\cos(qa). \tag{30}$$

Equations (29) and (30) are subject to: $\sum_i \alpha_i n_i = \bar{n} \leq l$, and $\sum_i \alpha_i = 1$, and to the assumption that the scenarios $sc_i$ are mutual strings scenarios. We assume that: $1 \ll \bar{n} \leq l$, so that the interaction between two strings is much larger than their kinetic energies $V(q) \gg \bar{e}(q)$. An important interaction is $V(2k_F)$, when $q = 2k_F = 2\pi a^{-1}\delta$, since this is the interaction that determines the pseudogap parameter $\Lambda$. The pseudogap parameter $\Lambda$ can easily be deduced from experiment, and from which one can get $V(2k_F)$. The reader should note, however, that $\Lambda$ is obtained by $V(2k_F)$ after averaging over the interactions with all the other strings, according to the Hartree-Fock diagrams.

4. THE SPECTRAL REPRESENTATION OF THE PSEUDOGAP STATE.

The ground state of the pseudogap symmetry is

$$|\Psi_0(t)>= \prod_{k_c,k_r=-a^{-1}\pi\delta}^{a^{-1}\pi\delta} |\Psi_{k_c}(t)>|\Psi_{k_r}(t)>. \tag{31}$$

In (31) $|\Psi_{k_c}(t)>$ is a ground state of a column string, and $|\Psi_{k_r}(t)>$ is a ground state of a row string. Hereafter we proceed only with column functions (without special notation), just for the pedagogical simplicity. The basic characteristics of the pseudogap state are given in the "Origin". Here we add the new feature which is the outcome of the wide spectral distribution of the string energy and the pseudogap parameter. Any ground state with specific wavenumber is a product of two such functions, one for each of the two anti-ferromagnetic states- A and B, as is shown in the "Origin". Consequently,

$$|\Psi_k(t)> = \frac{1}{\sqrt{2}}[|\Psi_k(t)>_A + |\Psi_k(t)>_B]. \tag{32}$$

The ground state and the excited states are defined by the following two field operators, $\gamma_{k,A}, \gamma_{k,B}$ - for particle-like operators, and $\eta_{k,A}, \eta_{k,B}$ - for anti-particle-like operators.

$$\gamma_{k,A}(t), \gamma_{k,B}(t) = \sum_i D_i(k,t)[\mp w_k C_k + v_k C_{\bar{k}}]_{A,B} \tag{33a}$$

$$\eta_{k,A}(t), \eta_{k,B}(t) = \sum_i D_i(k,t)[v_k C_k^\dagger \pm w_k C_{\bar{k}}^\dagger]_{A,B} \tag{33b}$$

In (33a) the minus, plus sign corresponds to $\gamma_{k,A}, \gamma_{k,B}$, respectively, and in (33b) the plus, minus sign corresponds to $\eta_{k,A}, \eta_{k,B}$, respectively. In (33a) and (33b), $D_i(k,t)$ is the spectral distribution of the field operators, which is given by

$$D_i(k,t) = d_i \exp[-iE_i(k)t] = d_i \exp[-i\sqrt{\varepsilon_i^2(k) + \Lambda_i^2(k)}\, t]. \tag{33c}$$

The operators $\gamma_{k,A}, \gamma_{k,B}$, and $\eta_{k,A}, \eta_{k,B}$ are defined as in the "origin" except that here they are weighted by their spectral distribution $D_i(k,t)$. This spectral distribution is the analog of $A_i(k,t)$ except that it is calculated by means of the two string interaction scenarios and its energy scale is at least an order of magnitude larger. This is so because, as shown in the "Origin", the basic excitation energies of the $\gamma_{k,A}^\dagger, \gamma_{k,B}^\dagger$, and $\eta_{k,A}^\dagger, \eta_{k,B}^\dagger$ are $E(k) = \sqrt{\varepsilon^2(k) + \Lambda^2(k)}$, where $\varepsilon(k) = [e(k) - e(k_F)] \ll \Lambda(k)$. Here it becomes $E_i(k) = \sqrt{\varepsilon_i^2(k) + \Lambda_i^2(k)} \simeq |\Lambda_i(k)|$, where $\varepsilon_i(k) = e_i(k) - e_i(k_F)$, and $-2k_F \leq k \leq 2k_F$. The field of the pseudogap state is given by

$$\psi_k(t) = \frac{1}{\sqrt{2}}[\begin{pmatrix}-w_k\\v_k\end{pmatrix}\gamma_{k,A}(t) + \begin{pmatrix}v_k\\w_k\end{pmatrix}\eta^\dagger_{k,A}(t) + \begin{pmatrix}w_k\\v_k\end{pmatrix}\gamma_{k,B}(t) + \begin{pmatrix}v_k\\-w_k\end{pmatrix}\eta^\dagger_{k,B}(t)]. \quad (34)$$

The Hamiltonian density for the pseudogap state is the field average of the energy operator, namely: $H_0 = i\psi^\dagger(x,t)\frac{d}{dt}\psi(x,t)$, where $\psi(x,t) = \sum_{k=-2k_F}^{2k_F}\psi_k(t)\exp(ikx)$. The range of $k$ in the sum is doubled because the pseudogap state is defined on both $k$ and $\bar{k}$. With the notion that $w_{\bar{k}}, v_{\bar{k}} = v_k, w_k$, and $\gamma_{\bar{k},A} = -\gamma_{k,A}$, $\eta_{\bar{k},B} = -\eta_{k,B}$, we can express the Hamiltonian as a sum over $|k| \leq k_F$, and get

$$H_0 = \frac{1}{2}\sum_{k=-k_F}^{k_F}\sum_i E_i(k)\{|d_{-i}|^2(1-2w_kv_k)\gamma^\dagger_{k,A}\gamma_{k,A} + |d_{+i}|^2(1+2w_kv_k)\eta^\dagger_{k,A}\eta_{k,A}$$

$$+|d_{+i}|^2(1+2w_kv_k)\gamma^\dagger_{k,B}\gamma_{k,B} + |d_{-i}|^2(1-2w_kv_k)\eta^\dagger_{k,B}\eta_{k,B} - 2\}. \quad (35)$$

In (35) the terms that are pre-factored by $\pm 2w_kv_k$ have been obtained by the product of $\psi^\dagger_{\bar{k}}(t)$ with $\frac{d}{dt}\psi_k(t)$. Note that there are excitation energies with the spectral components $E_{+i}(k) = E_i(k)(1+2w_kv_k)$, and energies with the spectral components $E_{-i}(k) = E_i(k)(1-2w_kv_k)$. The latter correspond to very low excitation energies, of order $\varepsilon_i^2(k)/2|\Lambda_i(k)|$, that are obtained from the difference between two very close energies. This made us assume for $E_{-i}(k)$ a new spectral distribution- $d_{-i}(k)$, which is roughly the difference between $d_i(k)$ and the very close distribution of $\Lambda_i(k)$. We have no knowledge about $d_{-i}(k)$, but we assume that $d_{-i}(k)$ takes substantial values at very small energies. However, the reader should differentiate between the energy- $E_{-i}(k)$, and its spectral distribution. When the distributions of $\varepsilon_i^2(k)$ and $|\Lambda_i(k)|$ overlap, the peak of the distribution of $\varepsilon_i^2(k)/2|\Lambda_i(k)|$ should move to higher energy than the peak of $\varepsilon_i^2(k)$, since the distribution of $|\Lambda_i(k)|$ is increasing around

the latter peak. The reader can have a rough evaluation of $d_{-i}(k)$ through the ARPES data that will be given in the end of this section. The distribution $d_{+i}(k)$ is roughly similar to $d_i(k)$, when the energy scale is doubled, because it is the distribution of $E_i(k) + \Lambda_i(k)$.

The Hamiltonian in the presentation of $\tilde{\psi}_{k,A,B} = \begin{pmatrix} C_k \\ C_{\bar{k}} \end{pmatrix}_{A,B}$ is given by

$$H_0 = \frac{1}{2} \sum_{k=-k_F}^{k_F} \sum_i |d_i|^2 E_i(k) \{ \tilde{\psi}_{k,A}^\dagger \begin{pmatrix} w_k^2 - v_k^2 - 2w_k v_k & -2w_k v_k \\ -2w_k v_k & v_k^2 - w_k^2 - 2w_k v_k \end{pmatrix} \tilde{\psi}_{k,A}$$

$$+ \tilde{\psi}_{k,B}^\dagger \begin{pmatrix} w_k^2 - v_k^2 + 2w_k v_k & 2w_k v_k \\ 2w_k v_k & v_k^2 - w_k^2 - 2w_k v_k \end{pmatrix} \tilde{\psi}_{k,B} \} \} \quad (36a)$$

With $\quad w_k^2, v_k^2 = \frac{1}{2}[1 \pm \frac{\varepsilon_i(k)}{E_i(k)}], \quad 2w_k v_k = -\left(\frac{\Lambda_i(k)}{E_i(k)}\right)_A, \quad$ and $\quad 2w_k v_k = \left(\frac{\Lambda_i(k)}{E_i(k)}\right)_B,$

equation (36a) becomes

$$H_0 = \frac{1}{2} \sum_{k=-k_F}^{k_F} \sum_i |d_i|^2 \{ \tilde{\psi}_{k,A}^\dagger [\varepsilon_i(k)\tau_3 + \Lambda_{i,A}(k)(I+\tau_1)] \tilde{\psi}_{k,A}$$

$$+ \tilde{\psi}_{k,B}^\dagger [\varepsilon_i(k)\tau_3 + \Lambda_{i,B}(k)(I+\tau_1)] \tilde{\psi}_{k,B} \}. \quad (36b)$$

The Hamiltonian in (36a,b) takes the form of (35) after diagonalization.

The Hamiltonian of (35) demonstrates clearly a separation into two bands, the low excitation energy band- $E_-(k) = E(k)(1 - 2w_k v_k)$, of $\gamma_{k,A}^\dagger$ and $\eta_{k,B}^\dagger$, and the high energy band- $E_+(k) = E(k)(1 + 2w_k v_k)$ of $\gamma_{k,B}^\dagger$ and $\eta_{k,A}^\dagger$. Due to the broad spectral distributions, the two bands may have some partial overlap. While the high energy band peaks around $2|\Lambda_k|$, the low energy band is gapless and enables the conductivity and the superconductivity of the system. This suggests that the field in (34) may also be divided into the low energy field $\psi_-$, and the high energy field $\psi_+$.

$$\psi_{k-}(t) = \frac{1}{\sqrt{2}} \sum_i [d_{-i} \begin{pmatrix} -w_k \\ v_k \end{pmatrix} \gamma_{k,A} \, e^{-iE_{-i}(k)t} + d_{-i}^\dagger \begin{pmatrix} v_k \\ -w_k \end{pmatrix} \eta_{k,B}^\dagger \, e^{iE_{-i}(k)t}] \quad (37a)$$

$$\psi_{k+}(t) = \frac{1}{\sqrt{2}} \sum_i [d_{+i} \begin{pmatrix} w_k \\ v_k \end{pmatrix} \gamma_{k,B} \, e^{-iE_{+i}(k)t} + d_{+i}^\dagger \begin{pmatrix} v_k \\ w_k \end{pmatrix} \eta_{k,A}^\dagger \, e^{iE_{+i}(k)t}] \quad (37b)$$

The propagators of the low energy band are

$$G_-(k,\omega) = \frac{1}{2} \sum_i |d_{-i}|^2 [\frac{1}{\omega - E_{-i}(k) + i\delta\omega} + \frac{1}{\omega + E_{-i}(k) + i\delta\omega}]. \quad (38a)$$

The propagators of the high energy band are

$$G_+(k,\omega) = \frac{1}{2} \sum_i |d_{+i}|^2 [\frac{1}{\omega - E_{+i}(k) + i\delta\omega} + \frac{1}{\omega + E_{+i}(k) + i\delta\omega}]. \quad (38b)$$

Equations (38a,b) present the propagators in a diagonal presentation. On the other hand, equation (37) in the "Origin" presents the propagator of the low energy band in a non-diagonal presentation, divided into its $(I, \tau_3, \tau_1)$ matrix components. The major difference between (38a) and the propagator of the "Origin" is the spectral distribution of the propagator here.

## 5. CONCLUDING REMARKS AND COMPARISON WITH EXPERIMENT.

The present paper is a continuation of a former paper by the present author, the "Origin" [11]. It accepts the basic perception of the "Origin", namely the aggregation of the holes into linear strings, as rows and columns that are arranged in checkerboard geometry. However, contrary to the "Origin", it presents a realistic model for the itinerancy of the strings, a model that results in a spectral distribution for the rates of the itinerancy. The basic features of this model also suggest a wide spectral distribution for the magnetic interaction between the strings. This is inferred from the nature of the propagation of the strings, but no quantitative analysis has been done. The wide spectral distribution of the interaction should result in a wide spectral distribution of the pseudogap parameter, which is a basic presumption in section 4. Apart from these wide distributions, all the basic assumptions of the "Origin" stay intact, as is shown above. Such assumptions are the existence of the anti-ferromagnetic phases A and B, the basic structure of the pseudogap ground state, the excitation operators, and the field. The ground state and the field are basically divided into two bands, the gapless low energy band, and the high energy band. Due to the wide distributions, the bands may be partially overlapped.

The ultimate test of any theory is its agreement with experiment. In the above discussions we have presented such agreements to several kinds of experiments, some qualitatively and some quantitatively. Generally speaking, the agreements with experiment that have been mentioned with respect to the "Origin" are still valid for the present paper, subject to some relevant broadening that is resulted from the obtained spectral distributions. As a reminder, these agreements are: 1) The UFS shown in Fig. 1b of the "Origin". 2) The 4x4 modulation by CDW that have been detected by many STM experiments [18-20]. 3) The incommensurate SDW that have been detected at the reciprocal vectors $\frac{\pi}{a}(1\pm 2\delta,1)$ and $\frac{\pi}{a}(1,1\pm 2\delta)$, by many elastic Neutron scattering experiments [21-25]. Another agreement to experiment is qualitative but essential. The present paper, as well as the "Origin", presents a unique model that yields a pseudogap and also a band that crosses the Fermi level, which enables conductivity and superconductivity.

Before testing comparison with other kinds of experiments, I wish to remark on a basic difficulty. Many relevant and important experiments in the field probe single electron or hole, whereas our theory deals with strings of holes. The present paper does not provide the translation of our results to experimental functions of single particle or anti-particle. This translation is obvious for Neutron elastic scattering, but less so for ARPES and electron tunneling. For the latter measurements we remark that the itineration rates of the strings are the same as the itineration rates of the individual hole within the string (or the individual spin that is the close neighbor to that hole).

The wide spectral nature of the magnetic interactions between the strings has resulted in wide spectra of the propagators in (38a,b). These wide spectra show up clearly in data which reflect density of states or spectroscopic intensities. Tunneling density of states in the normal pseudogap state has usually a finite minimum at zero energy, and rises continuously on both sides of the energy polarity, with no energy gap [31-33]. Elastic Neutron scattering measurements show incommensurate peaks at $\pm 2k_F$, and zero energy loss [21-25]. Energy Distribution Curves (EDC) of angular resolved photo-emission spectroscopy (ARPES) provide intensities of photo electrons as a function of energy and momentum. When the momentum is in the anti-nodal region, by the edge of the Brillouin-zone, the photo electrons are adjacent to the strings of holes, and reflect their states. There is much data of this sort, but here we discuss only a couple of papers that exhibit the two kinds of spectra $E_{-i}$ and $E_{+i}$. EDC of ARPES measurements on Bi2201 crystals on the anti-nodal direction show clearly the said two spectra superimposed [29]. The low energy spectrum of $E_{-i}$ is seen as a dispersion-less low energy shoulder in Fig. 2 and Fig. 4. The energy of the "shoulder" in Fig. 4M is roughly 20meV, but given that the claimed energy resolution is 10meV and given the uncertainty in the zero of the energy scale, the energy of this shoulder could be even smaller. The high energy peak in the same figures (the "hump") is roughly at 80meV. The difference between the shoulder energy and the high energy peak is roughly 60meV. In our model this should be equal to $2|\Lambda|$, so that for this material the pseudogap parameter equals 30meV.

The second paper which we would like to discuss reports ARPES on $YBa_2Cu_4O_8$ (YBCO124) [30]. All spectra were measured at 25K, when the sample is in the superconducting state. The measurements demonstrate asymmetry with respect to the direction of the Oxygen $O1$ chain- the Y-S direction ($(0,\pi)$ to $(\pi,\pi)$), versus the direction perpendicular to the $O1$ chain- the X-S direction ($(\pi,0)$ to $(\pi,\pi)$). This suggests that strings in the Y-S direction could correspond to higher doping level than strings in the X-S direction. The symmetry between columns and rows of holes in this material is broken. The authors remark that cleaved surfaces of samples of the YBCO family are known to exhibit over-doping qualities relative to the bulk. Therefore, results in the Y-S direction are suspect of being over-doped, and results in the X-S direction are supposed to better fit our under-doping requirement. High and low energy peaks were observed in both directions, and are shown in Figs. 1, 2, and 3 [30]. The low energy peak in the X-S direction goes down to the Fermi level. The high energy peak right at the symmetry point X is at 200meV, which suggest that $|\Lambda| \simeq 100 meV$, for this direction.